\documentclass{pasj00}

\begin{document}
\SetRunningHead{Saitoh et al.}{Shock-Induced Starburst During the First Encounter}
\Received{2008/4/28}
\Accepted{2009/1/7}

\title{Toward First-Principle Simulations of Galaxy Formation:\\
        II. Shock-Induced Starburst at a Collision Interface 
        During the First Encounter of Interacting Galaxies}

\author{Takayuki \textsc{R.Saitoh}$^1$,
        Hiroshi \textsc{Daisaka}$^2$,
        Eiichiro \textsc{Kokubo}$^{1,3}$,
        Junichiro \textsc{Makino}$^{1,3}$,
        Takashi \textsc{Okamoto}$^{4,5}$,\\
        Kohji \textsc{Tomisaka}$^{1,3}$,
        Keiichi \textsc{Wada}$^{1,3}$,
        Naoki \textsc{Yoshida}$^6$ (Project Milkyway)
}
\affil{$^1$ Center for Computational Astrophysics,
        National Astronomical Observatory of Japan, 2--21--1 Osawa,
        Mitaka-shi, Tokyo 181--8588}

\affil{$^2$ Graduate School of Commerce and Management, 
        Hitotsubashi University, Naka 2-1 
        Kunitachi-shi, Tokyo 186--8601}

\affil{$^3$ Division of Theoretical Astronomy,
        National Astronomical Observatory of Japan, 2--21--1 Osawa,
        Mitaka-shi, Tokyo 181--8588; 
        and School of Physical Sciences, Graduate University of Advanced Study (SOKENDAI)}

\affil{$^4$ Institute for Computational Cosmology, Department of Physics, 
        Durham University, South Road, Durham, DH1 3LE, UK}

\affil{$^5$ Center for Computational Sciences, University of Tsukuba
1-1-1,Tennodai, Tsukuba, Ibaraki 305-8577,Japan}

\affil{$^6$ Institute for the Physics and Mathematics of the Universe,
University of Tokyo, 5-1-5 Kashiwanoha, Kashiwa, Chiba 277-8568, Japan}

\email{saitoh.takayuki@nao.ac.jp,saitoh.takayuki@cfca.jp}


\maketitle

\begin{abstract}
We investigated the evolution of interacting disk galaxies using
high-resolution $N$-body/SPH simulations, taking into account the multiphase
nature of the interstellar medium (ISM).  In our high-resolution
simulations, a large-scale starburst occurred naturally at the collision
interface between two gas disks at the first encounter, resulting in the
formation of star clusters.  This is consistent with observations of
interacting galaxies.  The probability distribution function (PDF) of gas
density showed clear change during the galaxy-galaxy encounter.  The
compression of gas at the collision interface between the gas disks first
appears as an excess at $n_{\rm H} \sim 10~{\rm cm^{-3}}$ in the PDF, and
then the excess moves to higher densities ($n_{\rm H} \gtrsim 100~{\rm
cm^{-3}}$) in a few times $10^7$ years where starburst takes place.  After
the starburst, the PDF goes back to the quasi-steady state.  These results
give a simple picture of starburst phenomena in galaxy-galaxy encounters.
\end{abstract}

\section{Introduction}

Galaxy-galaxy interactions (collisions and mergers) are believed to play
important roles in shaping present-day galaxies, since they occur frequently
in the standard hierarchical model of galaxy formation (e.g.,
\cite{WhiteRees1978, LaceyCole1993, Kauffmann+1998, OkamotoNagashima2001}).
Interactions drive not only morphological transformations of galaxies but
also bursts of star formation (e.g., \cite{Young+1986, BarnesHernquist1992,
Smith+2007, CLi+2008}).

A number of researchers have studied roles of  galaxy-galaxy interactions
using numerical simulations (e.g., \cite{ToomreToomre1972, Hernquist1989,
MihosHernquist1994, MihosHernquist1996, BarnesHernquist1996, Springel2000,
Cox+2004, SpringelHernquist2005, Cox+2006}).  Many simulations incorporated
the interstellar medium (ISM) and models for star formation and supernova
(SN) feedback from massive stars.  In such simulations, the numerical
resolution is rather limited.  The typical gas mass resolution is about
$10^6~\Mo$.  Because of this limited mass resolution, it was impossible to
treat the formation of cold gas phase below $10^4~{\rm K}$.  (see \S 2 in
\cite{Saitoh+2008a}; hereafter paper I).  In most of early simulations, an
isothermal equation of state (EOS) of $10^4~{\rm K}$ was used for the ISM
(e.g., \cite{MihosHernquist1996}).  Even when the energy equation with a
cooling term was used,  the minimum temperature was set to be $\sim
10^4~{\rm K}$ (e.g., \cite{BarnesHernquist1996}).  Some authors
\citep{Springel2000, Cox+2004} adopted the EOS that became harder in higher
density.  They argued that such EOS mimicked the effect of turbulence
motions caused by SNe feedback from massive stars.  \citet{Struck1997}
employed a multiphase ISM model involving a low-temperature phase under
$10^4~{\rm K}$, although the modeling of ISM was rather simple and mass
resolution was low. 

Low-resolution simulations did not work well under the situation of strong
galaxy-galaxy interactions.  Observations of interacting galaxies often show
shock-induced, widespread star-formation activities involving star cluster
formation (e.g., the Antennae and the Mice galaxies), whereas previous
simulations showed that most of star formation activities was concentrated
to the central regions of the galaxies.  \citet{Barnes2004} demonstrated
that the collision simulations with the isothermal ISM and the
star-formation model governed by the local energy dissipation rate in shocks
can reproduce large-scale starburst formed at the collision interface during
the first collision.  However, his shock-induced star formation model
introduced an additional parameter and required the fine tuning of model
parameters for star formation.

The Schmidt law \citep{Schmidt1959} has been used as star formation model;
\begin{equation}
\frac{d \rho_{\rm *}}{dt} = C_* \frac{\rho_{\rm gas}}{t_{\rm dyn}},
\end{equation}
where $\rho_{\rm *}$ and $\rho_{\rm gas}$ are the stellar density and gas
density, respectively, $t_{\rm dyn} = (4 \pi G \rho_{\rm gas})^{-1/2}$ is
the local dynamical time, and $C_*$ is the dimensionless star formation
efficiency.  Since the isothermal EOS prevents ISM to form small-scale
structures, in most simulations, stars had to be formed in fairly
low-density gas.  Recent models employ the threshold density for star
formation, $n_{\rm H} \sim 0.1~{\rm cm^{-3}}$, above which star formation
occurs.  SNe releases energy into the surrounding ISM in the form of thermal
or kinetic energy, or internal turbulence motions.  When the value of $C_*$
was appropriately tuned, simulations based on these models successfully
reproduced observed global properties of star formation in local disk
galaxies, in particular the Schmidt-Kennicutt relation
\citep{Kennicutt1998}.  However, simulations with the isothermal EOS or the
temperature cutoff in cooling functions at $10^4~{\rm K}$ 
failed to reproduce three-dimensional structure of star formation
activities, simply because the star formation in these simulation took place
effectively everywhere in the gas disk.  In paper I, we demonstrated that
high-resolution simulation of the ISM with low temperature cooling below 
$10^4~{\rm K}$, combined with higher threshold density
for star formation, gave rise to thin and inhomogeneous star forming regions
as observed in isolated disk galaxies.  Our high-resolution simulations gave
a good description of small-scale structures of star forming regions, and at
the same time well reproduced global characteristics such as the
Schmidt-Kennicutt relation.

In this paper, we report the results of high-resolution simulations of a
collision between two disk galaxies with realistic ISM with cooling term and
high-density star-formation criterion for the Schmidt-type star formation
model.  This ISM model is based on the one used in paper I and it can
reproduce properties of star-formation in local disk galaxies, without the
fine tuning of the value of $C_*$.  Our main conclusion is that
high-resolution simulations can resolve shock-compressed regions, and, as a
result, these simulations can naturally reproduce a large-scale starburst at
the collision interface of gas disks during the first encounter.  Initially,
the density of the gas in the shock-compressed region reaches $n_{\rm H}
\sim 10~{\rm cm^{-3}}$.  This compressed gas collapses through radiative
cooling and gravitational instability in the timescale of $\sim 10^7~{\rm
yr}$ and reaches a high density at which star formation is triggered.
Thus, our modeling of the ISM, which requires sufficiently high-resolution,
naturally reproduces the shock-induced starburst.  A comparison run, which
was performed with the temperature cutoff in cooling function at $10^4~{\rm
K}$ and relatively low-density threshold for star-formation, displays only a
small enhancement of the star formation rate (SFR) during the first
encounter.  By using high-resolution simulations with a realistic star
formation model, we can successfully reproduce not only the quiescent star
formation in disk galaxies (paper I) but also the large-scale starburst at a
collision interface and nuclear starbursts observed in interacting galaxies
without any extra assumptions.

\section{Method and setup} \label{sec:methodology}

We used a parallel tree SPH code ASURA (Saitoh, in preparation) that
utilizes the special-purpose hardware GRAPE.  We first constructed
near-equilibrium, self-consistent galaxy models with a dark-matter halo and
an exponential disk, and then we converted a part of particles into gas
particles.  The $N$-body realization was generated by GalactICS
\citep{KuijkenDubinski1995}.  Our galaxy model is a low-mass, gas rich Sc
galaxy with the characteristic rotation velocity of $120~{\rm km~s^{-1}}$.
It has a gradually increasing rotation curve, which is consistent with
observations of nearby low-mass galaxies \citep{Sofue+1999}.  The
dark-matter halo mass and the disk mass are $1.05 \times 10^{11}~\Mo$ and
$6.3 \times 10^9~\Mo$, respectively.  The scale length of the exponential
disk is set to be $4~{\rm kpc}$ \citep{RobertsHaynes1994}.  We converted
$20~\%$ of the disk particles, $1.2 \times 10^9~\Mo$, into gas particles
with the exponential profile of an $8~{\rm kpc}$ scale length.  We also
converted $1~\%$ of the dark matter particles into gas particles with the
same profile as that of the dark matter halo.  The initial gas temperature
is set to be $10^4~{\rm K}$.  The galaxy-galaxy collision is characterized
by a prograde parabolic orbit with the pericentric distance $R_{\rm peri} =
7.5~{\rm kpc}$ and the initial separation of $75~{\rm kpc}$.  The relative
velocity at the moment of the pericenter passage is $\sim 370~{\rm
km~s^{-1}}$ for the point-mass approximation.  Two galaxies have the same
spin direction.  We prepare three initial models with different mass
resolutions.  Further details of the set up are described in Saitoh et al.
(in preparation).  The gravitational softening was set to be $20~{\rm pc}$
for all particles in all runs.  The particle mass and number of particles
are summarized in table \ref{tab:numbers}.

We use a cooling function for a gas with the metallicity of half the solar
value for the temperature range from $10~{\rm K}$ to $10^8~{\rm K}$
\citep{SpaansNorman1997, WadaNorman2001}.  A uniform heating from the
far-ultraviolet radiation is included.  The intensity of the far-ultraviolet
radiation is set to be the half of the solar neighborhood value
\citep{Wolfire1995}.

Models for the star formation and the SNe feedback are similar to those in
paper I.  We first adopt the model which allows a wide range of temperature
for ISM ($10 - 10^8~{\rm K}$) and uses the values $100~{\rm cm^{-3}}$ and
$100~{\rm K}$ as the threshold density and temperature for the
star-formation, respectively.  We add the label `C' in run names for these
models.  For comparison, we perform a run which only allows the ISM
temperature above $10^4~{\rm K}$ and employs $0.1~{\rm cm^{-3}}$ and
$15000~{\rm K}$ as the threshold density and temperature for the
star-formation, respectively.  The behavior of the ISM in this run should be
similar to those in previous simulations with isothermal ISM of $\sim
10^4~{\rm K}$.  The label `A' indicates this model.  We adopt $C_* = 0.033$
for all runs.

\begin{table*}
\begin{center}
\caption{Number and mass of particles for initial conditions of merger
simulations. The cutoff temperature for the ISM, threshold density and
temperature and efficiency for star formation are also
listed.}\label{tab:numbers}
\begin{tabular}{lccccccccc}
\hline
\hline
Run & ${N_{\rm DM}}^{\rm a}$ & ${N_{\rm Disk}}^{\rm b}$ & ${N_{\rm Gas}}^{\rm c}$ 
& $m^{\rm d}$ & ${T_{\rm cut}}^{\rm e}$ & ${n_{\rm th}}^{\rm f}$ & ${T_{\rm th}}^{\rm g}$ &${C_{*}}^{\rm h}$\\
\hline
M1A &  6~930~000 & 341~896 & 148~104 & $3 \times 10^4~\Mo$ & $10^4~{\rm K}$ &
$0.1~{\rm cm^{-3}}$ & $15000~{\rm K}$ & $0.033$ \\
M1C &  6~930~000 & 341~896 & 148~104 & $3 \times 10^4~\Mo$ & $10~{\rm K}$ &
$100~{\rm cm^{-3}}$ & $100~{\rm K}$ & $0.033$ \\
M2C & 13~860~000 & 683~678 & 296~322 & $1.5 \times 10^4~\Mo$ & $10~{\rm K}$ &
$100~{\rm cm^{-3}}$ & $100~{\rm K}$ & $0.033$ \\
M3C & 27~720~000 & 1~361~012 & 598~988 & $7.6 \times 10^3~\Mo$ & $10~{\rm K}$ &
$100~{\rm cm^{-3}}$ & $100~{\rm K}$ & $0.033$ \\
\hline
\end{tabular}\\
\end{center}
$^{\rm a}$The number of DM particles.
$^{\rm b}$The number of disk particles.
$^{\rm c}$The initial number of SPH particles.
$^{\rm d}$Mass of individual particles ($\Mo$).
$^{\rm e}$Cutoff temperature of cooling function (${\rm K}$).
$^{\rm f}$Threshold density of star formation (${\rm cm^{-3}}$).
$^{\rm g}$Threshold temperature of star formation (${\rm K}$).
$^{\rm h}$Star formation efficiency.
\end{table*}

\section{Starburst during the first encounter}

\begin{figure}[htb]
\begin{center}
\includegraphics[width = 0.47 \textwidth]{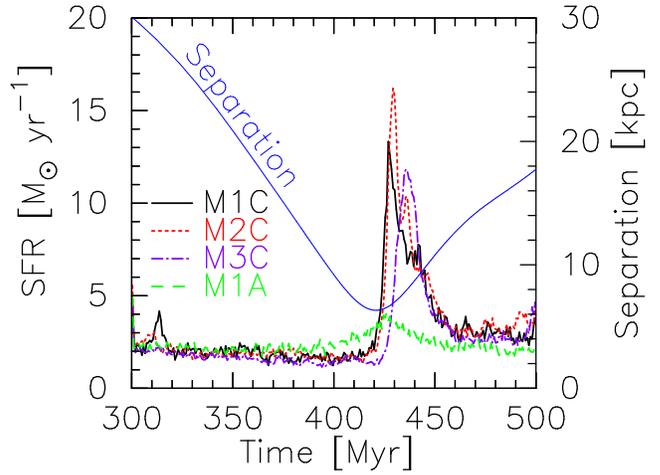}
\caption{Global SFR during the first encounter as a function of time.
Thick solid, dotted, dot-dashed, and dashed curves indicate the global SFR
for Runs M1C, M2C, M3C, and M1A, respectively.  The thin curve indicates the
separation between two galactic centers.
} \label{fig:sfr}
\end{center}
\end{figure}

Figure \ref{fig:sfr} shows global SFR during the first encounter.  Before
the encounter, each galaxy has a nearly constant SFR of $\sim 1~\Mo~{\rm
yr^{-1}}$.  For Runs M1C, M2C, and M3C, SFR increases by almost an order of
magnitude just after the pericenter passage ($\sim 430~{\rm Myr}$; see the
thin curve).  We call this peak as a {\it starburst}.  This duration of the
starburst is short and the phase of star formation returns quickly to the
quiescent star-formation phase after the first passage.  The slightly larger
SFR than those seen in the pre-collision might be explained by strong
non-axisymmertic perturbations from the companion galaxy
\citep{Noguchi1988}. The behavior of the starburst at the first encounter is
similar to that obtained by \citet{Barnes2004} who implemented a
shock-induced star formation.  Also, the difference between Runs M1C, M2C,
and M3C is very small.  We conclude that our result is independent of the
mass resolution as long as the resolution is sufficiently high. In contrast
to these `C' runs, Run M1A shows very small enhancement of SFR during the
first encounter.

The difference in SFR is owing to the hydrodynamic pressure of the gas that
forms a bulk of the giant filament.  The pressure of gas in Run M1A is
roughly a thousand times larger than those in Runs M1C, M2C, and M3C at the
same density, because the gas in Run M1A can not cool below $10^4~{\rm K}$
while those in Runs M1C, M2C, and M3C can cool down to $10~{\rm K}$.  In Run
M1A, the high pressure ISM in the contact interface resists collapse even
under significant compression.  As a result, the density of the ISM is not
increased significantly at the interface.  Since the timescale of the
pericenter passage of the galaxy-galaxy interaction is not very long (say,
$~10^7~{\rm years}$), in Run M1A, a smaller amount of the stars are formed
during the first encounter, as long as we adopt the Schmidt law for the
modeling of the star formation.

\begin{figure*}[htb]
\begin{center}
\includegraphics[width = 0.80 \textwidth]{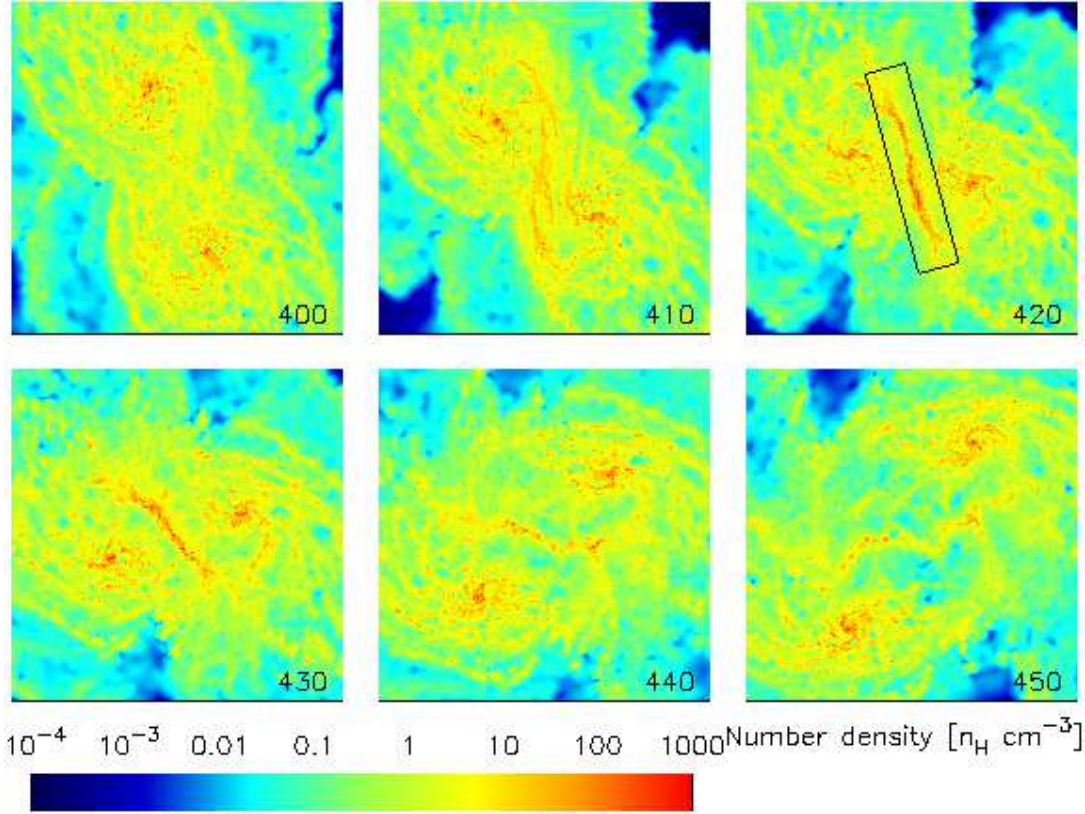}
\caption{Gas density distribution for Run M2C. Each panel shows a $16~{\rm
kpc} \times 16~{\rm kpc}$ region. The elapsed time in Myr is shown the
right-bottom corner of each panel. 
} \label{fig:Gas}
\end{center}
\end{figure*}

Figure \ref{fig:Gas} shows the gas density of Run M2C at the orbital plane
for six different epochs.  Two galaxies first follow the parabolic orbit
and then collide at the pericenter (top panels).  During the first
encounter, the maximum Mach number reaches $\sim 100$ for the cold ($T <
100~{\rm K}$) gas.  The ISM at the collision interface is strongly
compressed by shocks and forms a large and dense filament (see the top-right
panel of figure \ref{fig:Gas}; $t = 420~{\rm Myr}$).  We call it a {\it
giant filament}.  The length of this giant filament at its formation epoch
is $\sim 10~{\rm kpc}$.  We measured the gas mass in the rectangular box of
the size $10~{\rm kpc} \times 2~{\rm kpc} \times 2~{\rm kpc}$ along the
giant filament (the black rectangular box in the top-right panel of figure
\ref{fig:Gas}).  The gas mass of the giant filament is $\sim 9 \times
10^8~\Mo$.  This mass is about one quarter of the total gas mass in the
system.  The stellar mass formed during the starburst phase ($t =
420-450~{\rm Myr}$) becomes $\sim 20$ \% of the initial filament mass.  The
distribution of star forming regions is consistent with observations in
interacting galaxies (e.g., \cite{WhitmoreSchweizer1995}).  Overall
evolution of the colliding galaxies in the first encounter is quite
different from those in previous simulations with the isothermal ISM and the
Schmidt-type star formation models.  Our high-resolution simulation
naturally captures the shock-induced, large-scale starburst at the collision
interface.  A number of ``star clusters'' form in the giant filament at the
starburst phase.  We will study the formation of the star clusters in
further details in forthcoming papers.

\begin{figure}[htb]
\begin{center}
\includegraphics[width = 0.40 \textwidth]{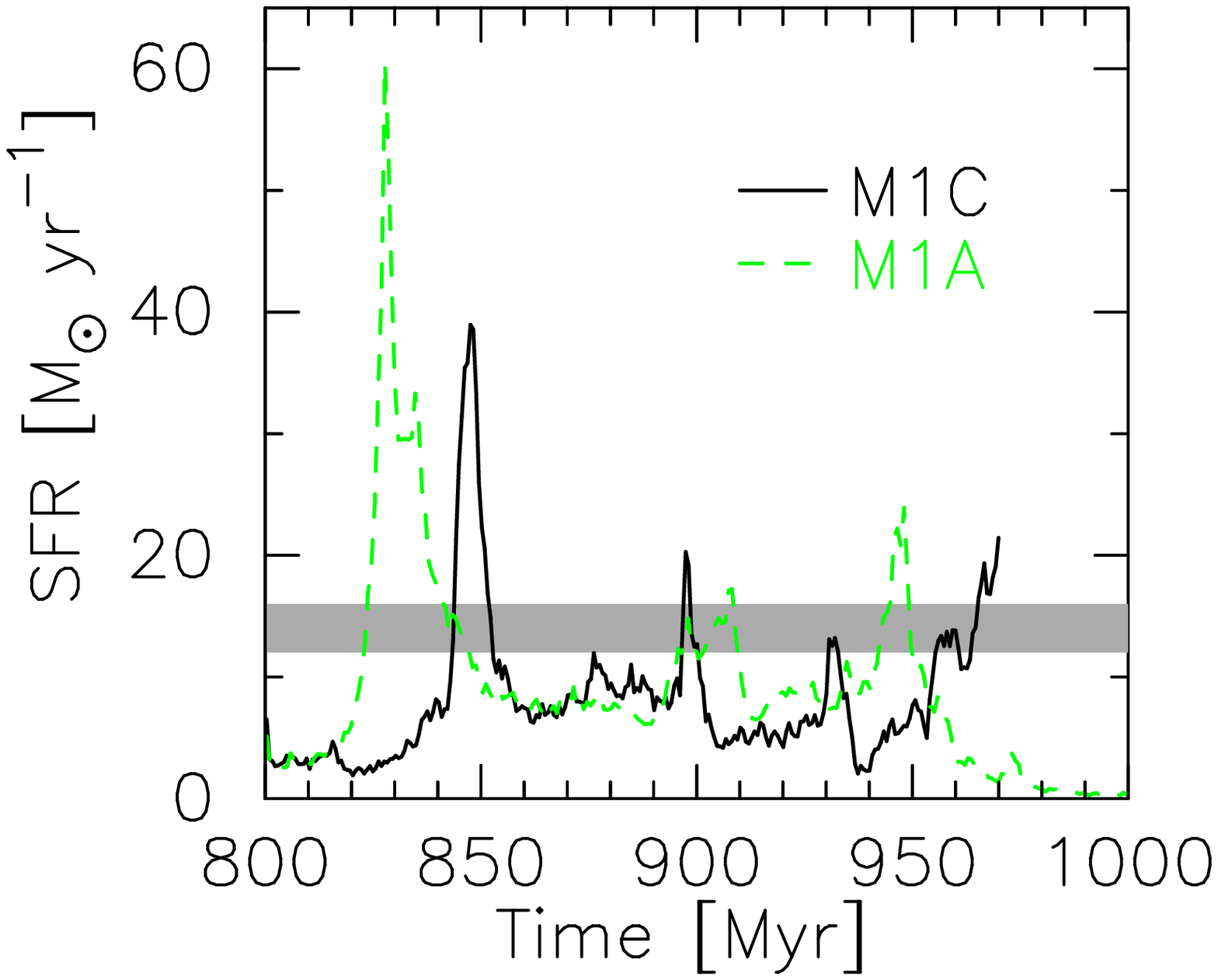}
\caption{Global SFR during the core-merger phase as a function of time.
Thick solid and dotted curves indicate the global SFR for Runs M1C and M1A,
respectively.  For reference, the peak SFRs at the first encounters is
$12-16~\Mo~{\rm yr^{-1}}$ (the horizontal gray zone), which is small to the
SFR peaks in the core-merger phase shown here.
} \label{fig:sfr_latephase}
\end{center}
\end{figure}

In addition to the shock-induced starburst in the giant filament, we also
find the starburst during a core-merging phase.  Figure
\ref{fig:sfr_latephase} follows the evolution of SFR during this phase. As
is shown in this figure, simulations with our model shows starbursts with
multiple peaks in the final core-merge phase. The star formation mainly
takes place within compact clumps in this phase. This is consistent with the
observations of ULIRGs \citep{SandersMirabel1996}.  Our simulations
successfully reproduce the nuclear starbursts as well as the large-scale
starbursts.  We further find the star formation enhancement in the tidal
tails.

\section{Probability distribution function during the starburst}

\begin{figure}[htb]
\begin{center}
\includegraphics[width = 0.45 \textwidth]{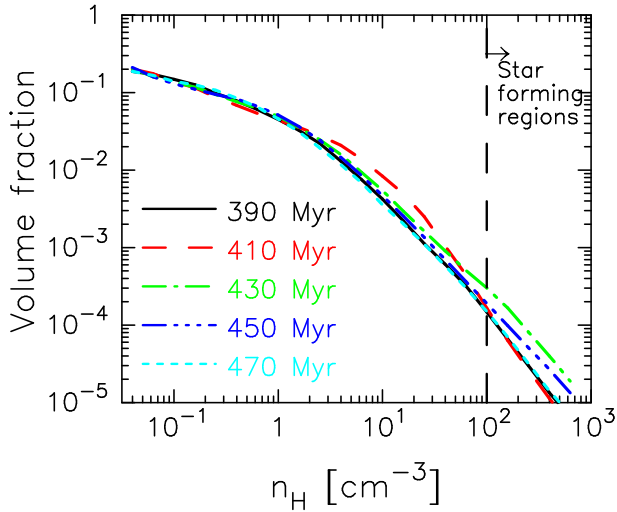}
\caption{Volume weighted probability distribution functions for the gas of
Run M2C for five different epochs.  The gas within $R < 10~{\rm kpc}$ and
$|Z| < 1~{\rm kpc}$, where $R$ is the distance from the mid-point of two
galaxies at the initial time in the orbital plane and $Z$ is the vertical
height from the orbital plane, is considered.
} \label{fig:pdf}
\end{center}
\end{figure}

\begin{figure}[htb]
\begin{center}
\includegraphics[width = 0.45 \textwidth]{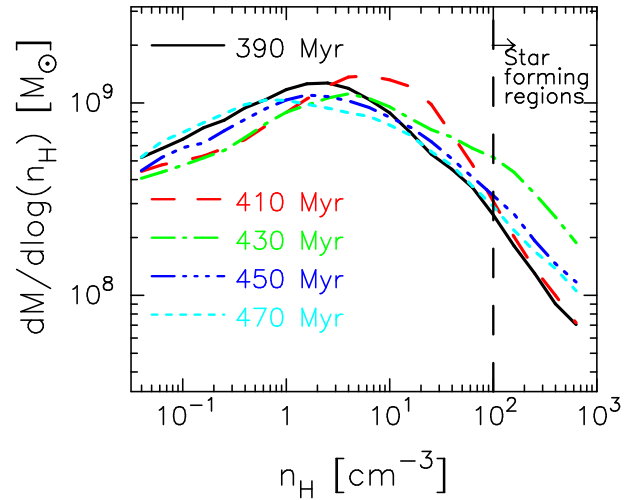}
\caption{Mass distribution as a function of the density of Run M2C for
five different epochs.  All gas in the system is included.
} \label{fig:mpdf}
\end{center}
\end{figure}

\begin{figure}[htb]
\begin{center}
\includegraphics[width = 0.45 \textwidth]{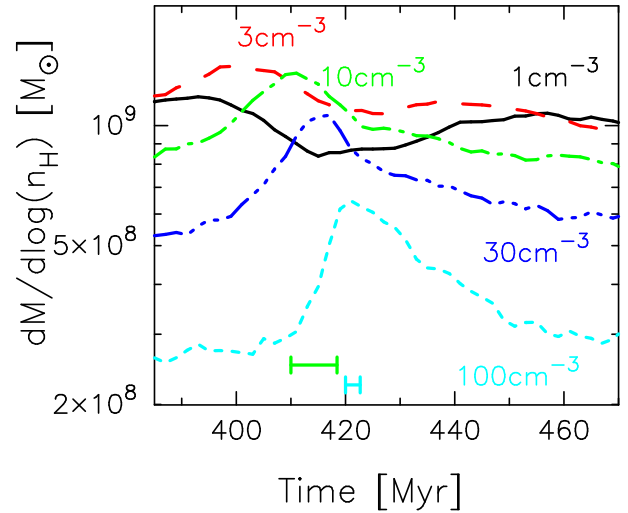}
\caption{Time evolution of gas mass for five different density ranges.  Long
(green) and short (light blue) horizontal bars correspond to local dynamical
times of densities at $n_{\rm H} = 10~{\rm cm^{-3}}$ and $100~{\rm
cm^{-3}}$, respectively.
} \label{fig:npdf}
\end{center}
\end{figure}

Figure \ref{fig:pdf} shows the volume weighted probability distribution
function (PDF) of Run M2C for several different epochs before, during, and
after the first encounter.  Although there are some noticeable changes in
the PDF with time, they are difficult to see because of the large vertical
scale.  In order to see the changes in the density distribution of the gas
more clearly, we plot the mass functions of the gas, $d M/d \log(n_{\rm
H})$, in figure \ref{fig:mpdf}.  Just before the encounter ($t = 410~{\rm
Myr}$), an excess of the gas (bump) appears at $n_{\rm H} \sim 10~{\rm
cm^{-3}}$.  By $t = 430~{\rm Myr}$, this bump moved to $n_{\rm H} \gtrsim
100~{\rm cm^{-3}}$, and by $t = 450~{\rm Myr}$ it almost vanished.  Note
that this period of $t = 430 - 450~{\rm Myr}$ coincides with the period of
the starburst.  Thus, we have a simple and clear picture of the
shock-induced starburst; Initially, large amounts of the gas is compressed
to $n_{\rm H} \sim 10~{\rm cm^{-3}}$.  Then, this compressed gas cools and
become denser through gravitational instability, and when the gas becomes
dense enough to form stars, the starburst takes place.  The starburst is led
mainly by star formation in massive clusters.  After the high-density gas is
consumed, the density PDF returns to the quasi-steady state.  In figure
\ref{fig:npdf}, we show the time variation of the mass in selected density
ranges.  The relative changes are larger for higher densities.  Moreover,
the peak is found at systematically later epochs for higher densities.  The
time delay from $n_{\rm H} = 10~{\rm cm^{-3}}$ to $100~{\rm cm^{-3}}$ is
$10~{\rm Myr}$.  This timescale corresponds to the local dynamical time,
$t_{\rm dyn}(n_{\rm H} = 10~{\rm cm^{-3}}) \sim 10~{\rm Myr}$, and it is
faster than the mean evolution timescale in isolated disk galaxies ($\sim
5~t_{\rm dyn}$; see figure 10 in paper I).  The lifetime of the mass excess
in each density range in figure \ref{fig:npdf} is around $20~{\rm Myr}$.  As
is expected, this timescale corresponds to that of the starburst.

\section{Summary} \label{sec:discussion}

We have studied the evolution of the multiphase ISM in the colliding
galaxies. Our simulations, for the first time, resolved the shock-induced
gas compression and starburst at the collision interface during the first
encounter of colliding galaxies, with the density dependent star formation
model (the Schmidt law). The behavior of the ISM is qualitatively different
from the results of previous studies with isothermal ISM.

Shock-induced starbursts, which have been observed in many interacting
galaxies, naturally take place in our high-resolution simulations without
any extra assumptions for the star-formation, such as the dependence on the
energy dissipation rate \citep{Barnes2004,Okamoto+2005}.  In previous
simulations with limited resolutions, significant shock-induced starbursts
took place only when the presence of the shock itself is used as the
criterion for star formation.

In paper I, we showed that high-resolution simulations with realistic ISM
and high-density star-formation criterion can reproduce the
Schmidt-Kennicutt relation without the need of the fine tuning of the value
of $C_*$.  Previous simulations with isothermal ISM and low-density
star-formation criterion required fine tuning.  In this paper, we showed
that high-resolution simulations can directly resolve shock-compressed
regions and these simulations can automatically reproduce shock-induced
starbursts.  High-resolution (particle mass less than $10^4~\Mo$) with a
realistic star formation model allows us to study both quiescent star
formation and starburst modes in $N$-body/SPH simulations of galaxy
formation.

\bigskip
We thank the anonymous referee for his/her insightful comments and
suggestions, which helped us to greatly improve our manuscript.  The author
(TRS) thanks Sachiko Onodera and Shinya Komugi for helpful discussions.
Numerical computations were carried out on Cray XT4/GRAPE system (project
ID:g07a19/g08a19) at CfCA of NAOJ.  This project is supported by
Grant-in-Aids for Scientific Research (17340059) of JSPS and Molecular-Based
New Computational Science Program of NINS.

\end{document}